\documentclass[aps,prapp,10pt,superscriptaddress,twocolumn]{revtex4-1}

\usepackage{tikz}
\usepackage[cmex10]{amsmath}
\usepackage{amssymb}
\usepackage{physics}
\usepackage{upgreek}
\usepackage{titlesec}
\usepackage{siunitx}
\usepackage{here}
\usepackage{tabularx}
\usepackage{xcolor}

\newcommand{\add}[1]{\textcolor{black}{#1}}
\newcommand{\erase}[1]{}

\usepackage{hyperref}
\hypersetup{
    colorlinks=true,
    linktocpage=true,
    linkcolor=blue,
    filecolor=magenta,      
    urlcolor=cyan,
    bookmarks=true,}
    
\graphicspath{{images/}}

\begin{document}
\title{Ultrastrong tunable coupler between superconducting LC resonators}

\author{T. Miyanaga}
\thanks{T.M. and A.T. are equally contributed to this work. \\ \href{mailto:akiyoshi.tomonaga@riken.jp}{akiyoshi.tomonaga@riken.jp}}
\affiliation{Department of Physics, Tokyo University of Science, 1--3 Kagurazaka, Shinjuku, Tokyo 162--0825, Japan}
\affiliation{RIKEN Center for Quantum Computing (RQC), 2--1 Hirosawa, Wako, Saitama 351--0198, Japan}

\author{A. Tomonaga}
\thanks{T.M. and A.T. are equally contributed to this work. \\ \href{mailto:akiyoshi.tomonaga@riken.jp}{akiyoshi.tomonaga@riken.jp}}
\affiliation{Department of Physics, Tokyo University of Science, 1--3 Kagurazaka, Shinjuku, Tokyo 162--0825, Japan}
\affiliation{RIKEN Center for Quantum Computing (RQC), 2--1 Hirosawa, Wako, Saitama 351--0198, Japan}

\author{H. Ito}
\affiliation{Department of Physics, Tokyo University of Science, 1--3 Kagurazaka, Shinjuku, Tokyo 162--0825, Japan}
\affiliation{RIKEN Center for Quantum Computing (RQC), 2--1 Hirosawa, Wako, Saitama 351--0198, Japan}

\author{H. Mukai}
\affiliation{Department of Physics, Tokyo University of Science, 1--3 Kagurazaka, Shinjuku, Tokyo 162--0825, Japan}
\affiliation{RIKEN Center for Quantum Computing (RQC), 2--1 Hirosawa, Wako, Saitama 351--0198, Japan}

\author{J. S. Tsai}\email{tsai@riken.jp}
\affiliation{Department of Physics, Tokyo University of Science, 1--3 Kagurazaka, Shinjuku, Tokyo 162--0825, Japan}
\affiliation{RIKEN Center for Quantum Computing (RQC), 2--1 Hirosawa, Wako, Saitama 351--0198, Japan}

\begin{abstract}
We investigate the ultrastrong tunable coupler for coupling of superconducting resonators.
Obtained coupling constant exceeds 1 GHz, and the wide range tunability is achieved both antiferromagnetics and ferromagnetics from $-1086$ MHz to 604 MHz.
The ultrastrong coupler is composed of rf-SQUID and dc-SQUID as tunable junctions, which connected to resonators via shared aluminum thin film meander lines enabling such a huge coupling constant. 
The spectrum of the coupler obviously shows the breaking of the rotating wave approximation, and our circuit model treating the Josephson junction as a tunable inductance reproduces the experimental results well.
The ultrastrong coupler is expected to be utilized in quantum annealing circuits and/or NISQ devices with dense connections between qubits. 
\end{abstract}

\maketitle

\section{Introduction}
The superconducting quantum circuit is one of the most outstanding platforms for quantum information processing and quantum sensing devices~\cite{Kwon2021Gate,Kjaergaard2020Superconducting,Degen2017Quantum}, and several important results have been reported in recent years~\cite{Arute2019Quantum,GQA2021Exponential,Wu2021Strong}.
These achievements are supported by the high degree of freedom in the design of Josephson junctions as nonlinear components and the fabrication technologies that have been developed in the field of semiconductors~\cite{Rosenberg20173D,Mukai2020Pseudo}.

However, major issues remain to be resolved before the realization of a practical universal quantum computer that implements quantum error correction~\cite{Alexeev2021Quantum}, and some of the devices \erase{that are expected to be feasible in the near future are}\add{such as} the quantum annealing machine~\cite{Kadowaki1998Quantum,Farhi2000Quantum} and noisy intermediate-scale quantum (NISQ) computer~\cite{Kandala2017Hardware,Preskill2018Quantum} \add{are expected to be feasible in the near future}.
These devices are application-specific circuits rather than general-purpose ones.
Thereby, when constructing complex qubit connections for a specific problem, circuit components require a high degree of design freedom.
For example, when superconducting circuit elements are placed on a wafer, nearest-neighbor interactions are commonly used; however, for the NISQ algorithm without error correction, full or partial full coupling is advantageous, as it allows a reducing in the depth of the quantum circuit (or entangling of many qubits at once) in order to finish the calculation within the coherence time~\cite{Zhou2020Quantum,Song2019Generation}.
In quantum annealing, practical problems such as circuit arrangement optimization and traveling salesman problems often have more complex interrelationships than the nearest-neighbor connection, and embedding these problems would require many more qubits in a sparsely coupled device~\cite{Zbinden2020Embedding}.

A flexible coupler enabling many qubits to be connected at once would be effective in such situations.
There are two approaches, using either a tunable or fixed coupler to connect circuit components, but the fixed coupling method is limited to applications in which the parasitic coupling can be neglected while keeping quantum states~\cite{Sheldon2016Procedure,Corcoles2015Demonstration} or when the coupled elements are treated as a single unit~\cite{Houck2012chip}.
The strategies for tuning the coupling strength are to control the effective energy exchange by tuning the energy of each component in time without a coupler (or with a fixed energy coupler) ~\cite{DiCarlo2009Demonstration, McKay2016Universal} or to control the tunable coupler in time~\cite{Harris2009Compound,Chen2014Qubit}.
For tunable couplers, the coupling energy is generally controlled by magnetic flux using a superconducting quantum interference device (SQUID)~\cite{Niskanen2007Quantum, Harris2009Compound, Weber2017Coherent, Sung2021Realization}.

Among the various types of coupler, using an ultrastrong inter-resonator coupler is one way to connect many qubits at once. 
By inter-resonator coupling, we expect circuit schemes that integrate full coupling circuits, such as in Ref.~\onlinecite{Mukai2019Superconducting}, or chimera graphs using a 10--20 qubits unit with full coupling, such as in Refs.~\onlinecite{Song201710,Song2019Generation}.
In these cases, the Hamiltonian of two qubits interacting through two resonators is written as
\begin{align}
 \mathcal{H}_\mathrm{2qr}/\hbar =  
        & \sum_{j=1,2} \qty[
        \frac{\omega_\mathrm{q}}{2}\hat{\sigma}_j^\mathrm{z} 
        +\omega_\mathrm{r}\qty(\!\hat{a}_j^\dagger \hat{a}_j+\frac{1}{2}\!) 
        +g_\mathrm{q} \hat{\sigma}_j^\mathrm{z} \qty(\!\hat{a}_j^\dagger +\hat{a}_j\!)
        ] \notag \\
        & -g_\mathrm{r}\qty(\hat{a}_1^\dagger - \hat{a}_1)\qty(\hat{a}_2^\dagger - \hat{a}_2) \,,
\label{eq:2QH}
\end{align}
where $\omega_\mathrm{q}$, $\omega_\mathrm{r}$, $g_\mathrm{q}$, and $g_\mathrm{r}$ represent the qubit frequency, the resonator frequency, the coupling constant between a qubit and a resonator, and the coupling constant between resonators. 
Also, $\hat{\sigma}_j^\mathrm{z}$ is the Pauli operator of qubits, and $\hat{a}_j$ and $\hat{a}{}^\dagger_j$ are the annihilation and creation operators for resonators, respectively.
The effective qubit-qubit coupling constant obtained from diagonalizing this Hamiltonian is $J_{12}=4g_\mathrm{r}g_\mathrm{q}^2/(\omega_\mathrm{r}^2-4g_\mathrm{r}^2)$~\cite{Billangeon2015Circuit}.
Usually, a superconducting quantum circuit is in the strong coupling regime ($g_\mathrm{q}/\omega_\mathrm{r}<0.1$).
In contrast, the ultrastrong coupling regime ($0.1\lesssim g_\mathrm{q}/\omega_\mathrm{r}<1$) and deep-strong coupling regime ($1\lesssim g_\mathrm{q}/\omega_\mathrm{r}$) require specific circuit designs~\cite{Yoshihara2016Superconducting,FriskKockum2019Ultrastrong}.
When the system is in the strong coupling regime, the effective qubit-qubit interaction $J_{12}$ is at least 100 times smaller than that in the case where qubits connect directly.
Thus, a inter-resonator coupler with a large coupling constant is required for a rapid two-qubit gate and quantum annealing for a full coupling circuit. 
The full coupling annealer in Ref.~\onlinecite{Mukai2019Superconducting} required the inter-resonator coupling constant $g_\mathrm{r}/2\pi=\pm400~\si{\mega\hertz}$ in addition to the deep-strong coupling between the qubit and the resonator.
However, among previous studies for inter-resonator coupling~\cite{Wulschner2016Tunable,Baust2015Tunable,Pierre2019Resonant}, Ref.~\onlinecite{Wulschner2016Tunable} reported that the coupling strength can be changed from $-320$ to 37 MHz.

In this work, we realize an inter-resonator ultrastrong coupler and achieve antiferromagnetic ($-1086$ MHz) to ferromagnetic (604 MHz) coupling tunability.
Our coupler consists of an rf-SQUID, which is connected to resonators via shared meander lines to gain a large mutual inductance. 
This could enables the simultaneous coupling of multiple qubits for quantum annealing and NISQ devices~\cite{Mukai2019Superconducting}.
In terms of circuit quantum electrodynamics, our coupler exceeds $g_\mathrm{r}/\omega_\mathrm{r}>0.1$, which can be called an ultrastrong coupling regime between resonators~\cite{Niemczyk2010Circuit}
Furthermore, we observe the breaking of the rotational wave approximation in the spectrum measurement.
We also confirm that the coupler can be turned on and off by comparing the simulation and measurement results, including the crosstalk between the input and output ports.
Also, inter-resonator couplings are expected to be applied for \add{a scalable} quantum computation~\cite{stassi_scalable_2020}\erase{using}, a cat code~\cite{Wang2016Schrodinger}, \erase{and }a holonomic gate~\cite{Wang2016Holonomic}, a parametric amplifier~\cite{Tian2008Parametric}, \add{and} a beam-splitter~\cite{Zheng2017Nonreciprocal}.

\section{Physical system}\label{sec1}
\begin{figure}[h!]
\centering
\includegraphics[width=85mm]{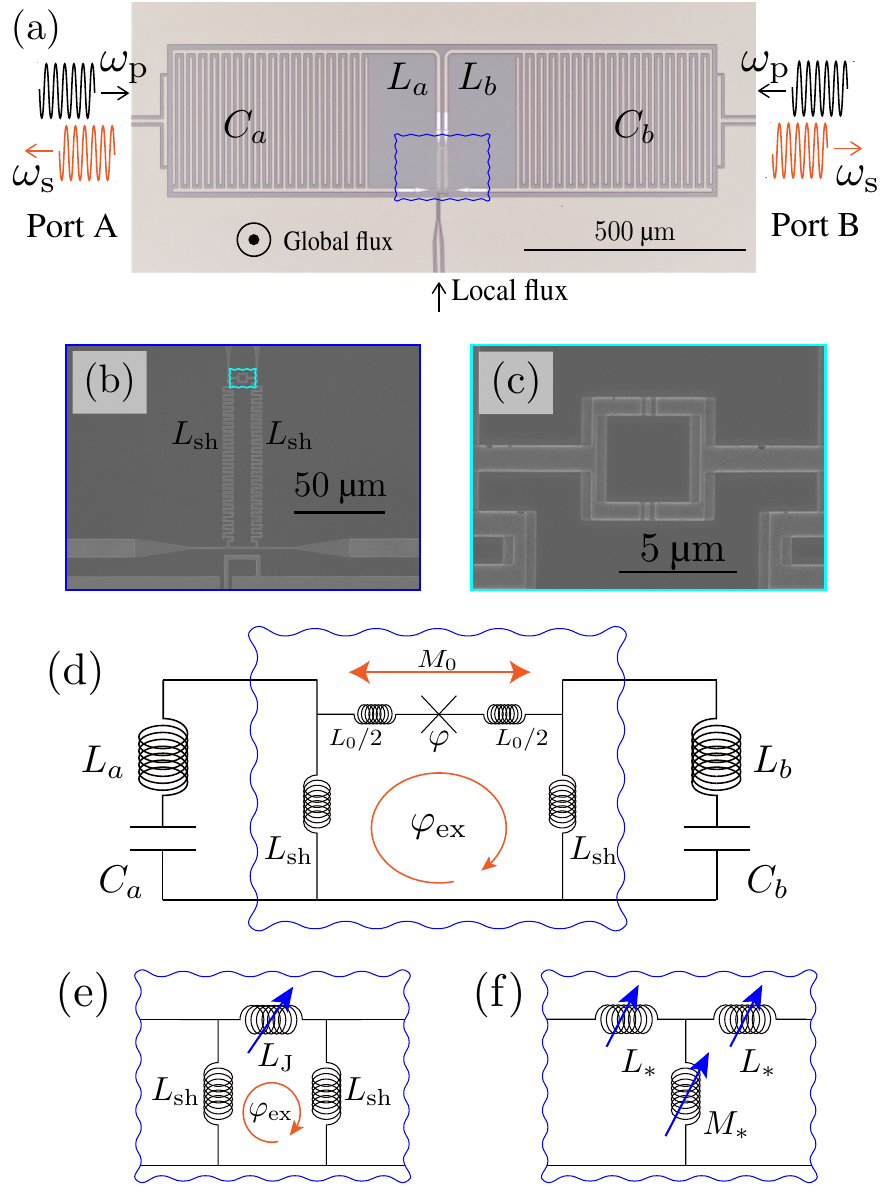}
    \caption{
    (a) Optical microscope image of the coupling circuit with rf-SQUID, fabricated from 50~nm niobium thin film on a high-resistance silicon wafer.
    Resonators A and B have the same design, but the frequency of resonator A is lower than that of resonator B, as seen from the reflection measurement described in section \ref{sec:cross} owing to the fabrication error or the asymmetrically designed local flux line.
    The sample holder has a coil to bias a uniform magnetic field from the back face of the sample.
    Input and output ports (ports A and B) are ``T''-shape conductors at the right and left ends.
    $\omega_p$ and $\omega_s$ represent the probe and signal microwave tones from the circuit.
    (b) SEM image of coupler (rf-SQUID) fabricated from aluminum film by double-angle shadow evaporation with Josephson junctions. $L_\mathrm{sh}$ represents the shared inductance between the resonator and the coupler formed by the aluminum meander line.
    (c) Enlarged image of the dc-SQUID area.
    (d) Circuit diagram of the system. The dc-SQUID is depicted as the single junction ``$\times$''.
    $\varphi_\mathrm{ex}$ can be changed by the global flux and local flux line, and the magnetic flux for the dc-SQUID is changed only by the global flux. 
    $M_0$ is the geometric mutual inductance and $L_0$ is the unshared loop inductance of the rf-SQUID.
    (e) Equivalent circuit of (d) and the junction is drawn as the classical tunable inductance $L_\mathrm{J}$.
    (f) Equivalent circuit of (e) in star-delta transformation.
    }
\label{Design1Circuit}
\end{figure}
The circuit model (Hamiltonian) of the system with two resonators connected by the tunable coupler is derived from circuit equations.
Figure~\ref{Design1Circuit}(a) shows an optical microscope image of the fabricated circuit, where resonators consist of interdigital capacitors and thin-film line inductors.
The rf-SQUID as a coupler, shown in Fig.~\ref{Design1Circuit}(b), is galvanically connected to resonators via meander lines and the dc-SQUID [Fig.~\ref{Design1Circuit}(c)] improves the tunability of a coupler to change the effective junction inductance of the rf-SQUID.
The transition frequency from the ground state to the first excited state of the rf-SQUID is designed to be sufficiently larger than the frequency of resonators, so that the rf-SQUID is always in the ground state and can be treated as a classical tunable inductance~\cite{Harris2009Compound}.
Thus, the term of the rf-SQUID is not explicitly represented in the following calculations of the Hamiltonian.
From the above treatment, the system can be described as the circuit diagram shown in Fig.~\ref{Design1Circuit}(d).
When we consider the dc-SQUID in Fig.~\ref{Design1Circuit}(c) as a tunable Josephson junction with inductance $L_\mathrm{J}(\varphi)=L_\mathrm{J0}/\cos\varphi + L_0$, which depends on the phase difference between the two ends of the junction, the circuit of the coupler can be described as in Fig.~\ref{Design1Circuit}(e) with external flux bias.
Considering that the rf-SQUID is always in the ground state, $\varphi$ is determined to minimize the potential of the rf-SQUID,
\begin{align}
    \mathcal{U}=\qty(\frac{\Phi_0}{2\pi})^2\qty[\frac{1}{2}\frac{\qty(\varphi-\varphi_\mathrm{ex})^2}{2L_\mathrm{sh}+L_0}-\frac{\cos{\varphi}}{L_\mathrm{J0}}]\,,
\end{align}
where $\Phi_0$ is the flux quantum.

Using the star-delta $(Y-\Delta)$ transformation~\cite{Tian2008Parametric,Mariantoni2008Two}, we can deal with the coupler as the inductance modulation $L_*\add{(\varphi_\mathrm{ex})}$ of resonators and the mutual inductance $M_*\add{(\varphi_\mathrm{ex})}$ between resonators, as shown in Fig.~\ref{Design1Circuit}(f), where
\begin{align}
 M_\mathrm{*}\add{(\varphi_\mathrm{ex})}&=\frac{L_\mathrm{sh}^2}{2L_\mathrm{sh}+L_\mathrm{J}} + M_0 \,, \label{eq:Mstar}\\
 L_\mathrm{*}\add{(\varphi_\mathrm{ex})}&=\frac{L_\mathrm{sh}L_\mathrm{J}}{2L_\mathrm{sh}+L_\mathrm{J}}\,.
 \label{eq:Lstar}
\end{align}
Then, the Lagrangean for this circuit is derived as
\begin{align}
 \mathcal{L}_\mathrm{2r}=
 &\frac{1}{2}\qty(L_a+L_*)I_a^2
 +\frac{1}{2}\qty(L_b+L_*)I_b^2 \notag \\
 &+\frac{1}{2}M_*(I_a-I_b)^2 
 -\frac{Q_a^2}{2C_a}-\frac{Q_b^2}{2C_b} \\
 =&\frac{1}{2}L'_a{\dot{Q}_{\add{a}\erase{1}}}^2
 +\frac{1}{2}L'_b{\dot{Q}_{\add{b}\erase{2}}}^2-M_*\dot{Q}_{\add{a}\erase{1}}\dot{Q}_{\add{b}\erase{2}}
 -\frac{Q_a^2}{2C_a}-\frac{Q_b^2}{2C_b}\,, 
\label{eq:L2r}
\end{align}
where $Q_k$ is the charge at $C_k$, the resonator current $\dot{Q}_k=I_k$, the effective inductance of resonators $L'_k\add{(\varphi_\mathrm{ex})}=\add{L_k+L_*}\erase{L_{k0}}+M_*$\erase{, and the bare inductance of resonators $L_{k0}\add{(\varphi_\mathrm{ex})}=L_k+L_*$} $(k \in \qty{a,b}\,)$.
When $Q_k$ is taken to be the canonical coordinates, the conjugate momentum $P_k$ can be written as
\begin{align}
 P_a\equiv\frac{\partial\mathcal{L}_\mathrm{2r}}{\partial \dot{Q}_a}
 &=L'_a\dot{Q}_a - M_*\dot{Q}_b \,,\\
 P_b&\equiv L'_b\dot{Q}_b - M_*\dot{Q}_a 
 \,.
\label{eq:pq}
\end{align}
Consequently, we obtain the Hamiltonian of the whole circuit as
\begin{align}
 \mathcal{H}_\mathrm{2r}&=\sum_{k=a,b} P_k\dot{Q}_k-\mathcal{L}_\mathrm{2r} \\
 &=\frac{P^2_a}{2L_{\mathrm{m}a}}+\frac{P^2_b}{2L_{\mathrm{m}b}}
 +\frac{P_a P_b}{M_\mathrm{m}}
 +\frac{Q_a^2}{2C_a}+\frac{Q_b^2}{2C_b}\,,
\label{eq:H2r0}
\end{align}
where the effective mass of each resonator $L_{\mathrm{m}a}\add{(\varphi_\mathrm{ex})}\equiv L'_a-M_*^2/L'_b$, $L_{\mathrm{m}b}\add{(\varphi_\mathrm{ex})}\equiv L'_b-M_*^2/L'_a$ and effective coupling mass $M_\mathrm{m}\add{(\varphi_\mathrm{ex})}\equiv -M_*+L'_a L'_b/M_*$.

Moreover, to quantize the circuit, we define annihilation and creation operators $\hat{a}$, $\hat{a}^\dagger$, $\hat{b}$, and $\hat{b}^\dagger$ 
($\hat{k}=\qty(Z_k Q_k + i  P_k)/\sqrt{2\hbar Z_k}$ and its Helmert conjugate), where characteristic impedance $Z_k \equiv \sqrt{L_{\mathrm{m}k}/C_k}$ and $\comm{Q_k}{P_l}=i\hbar\delta_{kl}$ ($k,l \in \qty{a,b}$).
Then, Hamiltonian Eq.~\eqref{eq:H2r0} can be written as
\begin{align}
 \mathcal{H}_\mathrm{2r}/\hbar =
    \omega_a\qty(\!\hat{a}^\dagger \hat{a}+\frac{1}{2}\!) 
 & + \omega_b\qty(\!\hat{b}^\dagger \hat{b}+\frac{1}{2}\!) \! \notag \\
 & -g_\mathrm{r}\qty(\hat{a}^\dagger\!-\hat{a})\qty(\hat{b}^\dagger\!-\hat{b})\,,
\label{eq:H2r}
\end{align}
where the \erase{resonant}\add{resonator} frequency $\omega_k\add{(\varphi_\mathrm{ex})}$ and the coupling constant $g_\mathrm{r}\add{(\varphi_\mathrm{ex})}$ are defined as
\begin{align}
    \omega_k\add{(\varphi_\mathrm{ex})}&\equiv1/\sqrt{L_{\mathrm{m}k}C_k}\,, \label{eq:wk}\\
    g_\mathrm{r}\add{(\varphi_\mathrm{ex})}&\equiv \sqrt{Z_a Z_b}/(2M_\mathrm{m}) \,.
    \label{eq:gr}
\end{align}
\add{Since the resonator is connected to SQUID, these dressed frequencies depend on external flux bias $\varphi_\mathrm{ex}$. As discussed later (especially in section \ref{sec:cross}), we define the bare resonator frequency $\omega_{k0}$ as a specific $\omega_k$ at a point of $g_\mathrm{r}=0$ ($M_*=0$).
However, this bare frequency does not correspond to $1/\sqrt{L_kC_k}$ due to the existence of $L_*$ (galvanically coupled to SQUID).
}
\erase{This frequency and coupling constant depend on the magnetic field applied to the rf-SQUID and dc-SQUID.}
If $M_*$ is sufficiently smaller than $\add{L_{k}+L_*}\erase{L_{k0}}$, the coupling constant can be expressed as $\hbar g_\mathrm{r}\simeq M_* I_{\mathrm{zpf}a}I_{\mathrm{zpf}b}$ using the zero point fluctuation current of \erase{bare }resonators $I_{\mathrm{zpf}k}=\sqrt{\hbar\add{\tilde{\omega}_k}\erase{\omega_{k0}}/2\add{(L_k+L_*)}\erase{L_{k0}}}$, where the \add{approximated} \erase{bare resonant }frequency $\add{\tilde{\omega}_{k}(\varphi_\mathrm{ex})}\erase{\omega_{k0}}=1/\sqrt{\add{(L_k+L_*)}\erase{L_{k0}}C_k}$.

Hamiltonian Eq.~\eqref{eq:H2r} can be exactly diagonalized by the Bogolivbov transformation $\hat{c}_\pm=u_a^{(\pm)} \hat{a} + v_a^{(\pm)} \hat{a}^\dagger+u_b^{(\pm)} \hat{b} + v_b^{(\pm)} \hat{b}^\dagger~$.
Then, the two harmonic modes are obtained as
\begin{align}
    \omega_\pm^2&=\frac{\omega_a^2+\omega_b^2
    \pm\sqrt{(\omega_a^2-\omega_b^2)^2+16g_\mathrm{r}^2\omega_a\omega_b}}{2}
    \,,
    \label{eq:wpm}
\end{align}
where $\{u_k^{(\pm)}, v_k^{(\pm)}\}\in \mathbb{C}$ and $\comm{\hat{c}_s}{\hat{c}_t}=i\hbar\delta_{st}$ (${s,t}\in \{+,-\}$).
In the two-resonator-coupled system, these eigenmodes $\omega_\pm\add{(\varphi_\mathrm{ex})}$ are observable, but not each resonator frequency $\omega_k\add{(\varphi_\mathrm{ex})}$~\cite{Billangeon2015Circuit} (see \ref{sec:A0}).
In addition, from Eq.~\eqref{eq:wpm}, $\omega_-$ becomes imaginary when $g_\mathrm{r}$ exceeds $\sqrt{\omega_a\omega_b}/2$, but taking into account a real circuit, such as in Fig.~\ref{Design1Circuit}(f), $g_\mathrm{r}$ is limited by the mutual inductance.
When we take the limit of the mutual inductance $M_*$, the coefficients of Eq.~\eqref{eq:H2r} converge to
\begin{align}
\lim_{M_* \to \infty}\omega_k &=
    1/\sqrt{(L_{a0}+L_{b0})C_k}\equiv\omega_k^\mathrm{inf}\,, \\
\lim_{M_* \to \infty} g_\mathrm{r} &=
    \sqrt{\omega_a^\mathrm{inf}\omega_b^\mathrm{inf}}/2 \,.
    \label{eq:limit}
\end{align}
In terms of a physical aspect, when $M_*$ is sufficiently large, electric current can no longer flow in $M_*$.
Then, $\omega_{-}=0$ and $\omega_{+}$ is an LC resonance arising from the combined two inductors and capacitors in the circuit, where $I_a=I_b$ in Eq.~\eqref{eq:L2r}.

\section{Measurement results}
\begin{figure*}[t!]
\centering
\includegraphics[width=180mm]{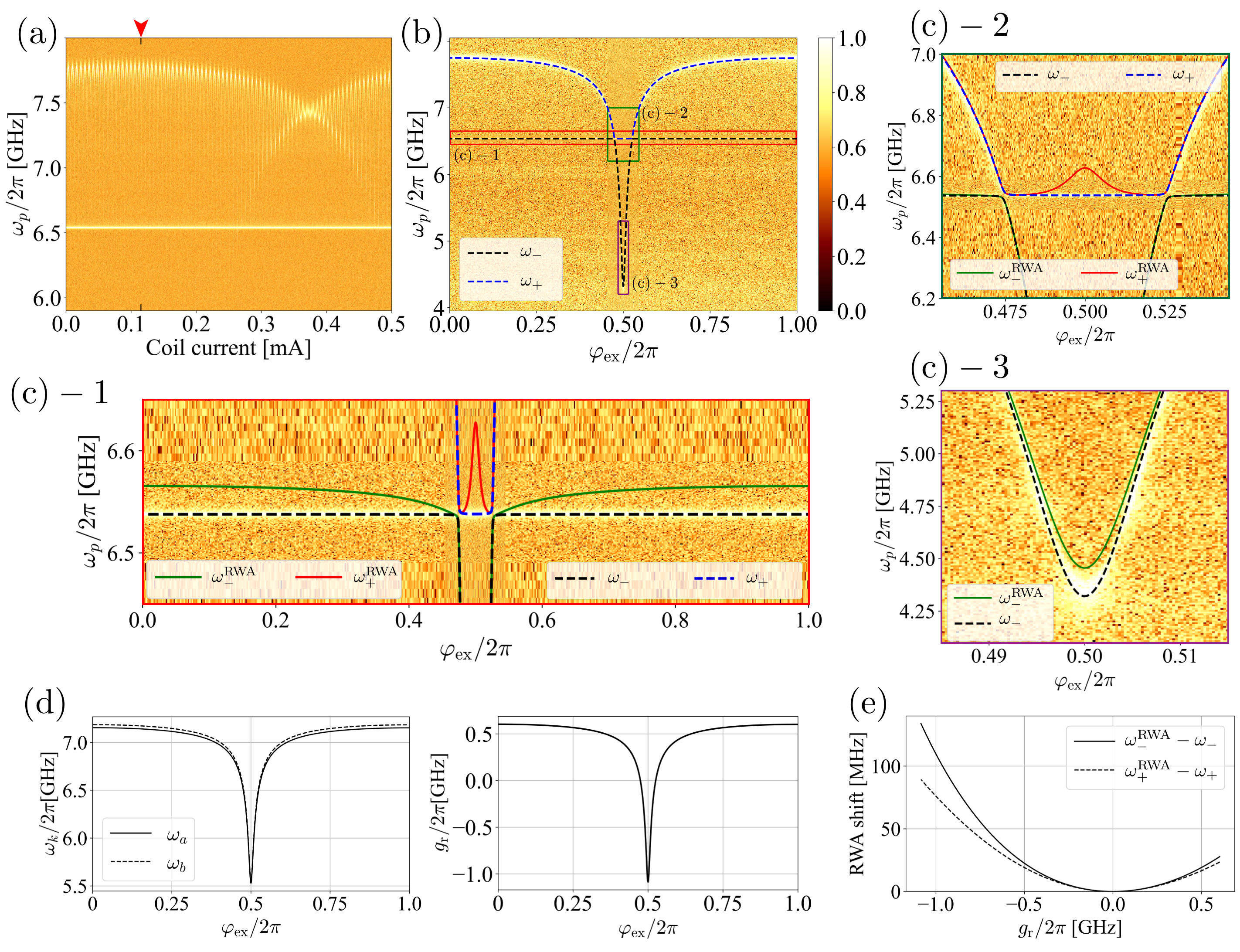}
    \caption{
    (a) Measured spectrum of the sample in Fig.~\ref{Design1Circuit} plotted against the coil current.
    \add{The background (off-resonant point) transmission signal amplitude is around 0.5 due to the crosstalk between input and output ports as shown in Fig~\ref{Chip} (\ref{sec:A})}
    (b) Measured spectrum plotted against $\varphi_\mathrm{ex}$ as local flux bias at the fixed coil current of 115~$\si{\micro}$A represented in (a) by the red arrow. Dashed lines are fitted curves using the eigenvalues of Hamiltonian Eq.~\eqref{eq:H2r} with the parameters shown in Table~\ref{table:SampleData}.
    (c)-1 (red), (c)-2 (green) and (c)-3 (purple) Enlarged images of several regions in (b). Dashed lines are the same as in (b) and solid lines represent the eigenvalues of Hamiltonian with the rotating wave approximation [Eq.~\eqref{eq:HRWA}] using the same parameters in Table~\ref{table:SampleData}.
    (d) Coefficients of the Hamiltonian Eq.~\eqref{eq:H2r} plotted against the flux bias obtained from the fitting in (b).
    (e) Difference between dashed lines ($\omega_\pm$) and dashed lines ($\omega_\pm^\mathrm{RWA}$) in (b) and (c) as a function of coupling constant.
    }
\label{Design1Spectrum}
\end{figure*}
\begin{table}[t]
    \caption{List of the fitted parameters in Fig.~\ref{Design1Spectrum}. 
    In the fitting function, we take into account $M_0$ and $L_0$ in Eqs.~\eqref{eq:Mstar} \add{and \eqref{eq:Lstar}} and also the offset of inductance modulation $\gamma$ in $L_\mathrm{J}(\varphi)=L_\mathrm{J0}/\qty(\cos\varphi-\gamma)+L_0$ (see \ref{sec:B}). \add{Bare resonator frequencies $\omega_{a0}=6.522$~GHz and $\omega_{b0}=6.551$~GHz are also obtained from this fitting parameters.}
    }
    \centering
    \label{table:SampleData}
  \begin{tabularx}{0.48\textwidth}{Xccc}
    \hline
    Name        & \quad  Symbol \quad\quad & Value \quad   &  \, Unit \quad\\ 
    \hline \hline
Self-inductance              & $L_{a,b}$        &  2.023  &  $\si{\nano\henry}$    \\
Capacitance of resonator A      & $C_{a}$          &  184.3  &  $\si{\femto\farad}$    \\
Capacitance of resonator B      & $C_{b}$          &  182.7  &  $\si{\femto\farad}$    \\
Shared inductance            & $L_\mathrm{sh}$  &  0.446  &  $\si{\nano\henry}$    \\
Junction inductance          & $L_\mathrm{J0}$  &  1.210  &  $\si{\nano\henry}$    \\
Geometric\phantom{a}mutual\phantom{a}inductance  & $M_\mathrm{0}$   &  0.381  &  $\si{\nano\henry}$    \\
Unshared loop inductance     & $L_\mathrm{0}$   &  0.177  &  $\si{\nano\henry}$    \\
Offset\phantom{a}of\phantom{a}inductance\phantom{a}modulation & $\gamma$         &  0.053  &  -   \\
    \hline
\end{tabularx}
\end{table}
Figure~\ref{Design1Spectrum} shows the measured spectrum of the transmission signal from ports A to B in the circuit in Fig.~\ref{Design1Circuit}.
Since the area size of the dc-SQUID loop is designed to be 100 times smaller than that of the rf-SQUID, Fig.~\ref{Design1Spectrum}(a) shows that the energy of the rf-SQUID changes 100 times while the inductance of the junction (dc-SQUID) modulates one period.
Figure~\ref{Design1Spectrum}(b) shows the spectrum measured by applying magnetic flux from the local flux line when the coil current is biased where the inductance of the dc-SQUID is around minimized (see \ref{sec:A}).
\add{The spectrum shows two eigenmodes $\omega_\pm$ [Eq.~\eqref{eq:wpm}], which consist of in-phase (parallel) and out-of-phase (anti-parallel) modes for two resonators~\cite{baust_ultrastrong_2016}. The in-phase mode does not affect the SQUID loop current and is independence of the flux bias.
These two modes are replaced each other between $\omega_\pm$ at $g_\mathrm{r}=0$ (see \ref{sec:A0}).}
Dashed lines are the result of fitting using \add{eigenmodes of} Hamiltonian Eq.~\eqref{eq:H2r} with the values of the circuit elements in Table~\ref{table:SampleData}; the physical system is reproduced well.
\add{The raw data of the spectrum without fitting curves are shown in Fig~\ref{Raw} (see \ref{sec:B})}
The coefficients of Hamiltonian Eq.~\eqref{eq:H2r} obtained from the fitting are shown in Fig.~\ref{Design1Spectrum}(d), and the coupling constant $g_\mathrm{r}/2\pi$ can be changed in the range from -1086 to $604~\si{\mega\hertz}$.
This coupling strength and tunability satisfy the $\pm 400~\si{\mega\hertz}$ range suggested in Ref.~\onlinecite{Mukai2019Superconducting}.
In addition, from Fig.~\ref{Design1Spectrum}(d), the ratio of the coupling constant to resonator frequency $\abs{g_\mathrm{r}}/\max\qty{\omega_a, \omega_b}\simeq 0.20$ at $\varphi_\mathrm{ex}/2\pi=0.5$ exceeds 0.1; thus, we can consider that our coupler is in the ultrastrong coupling regime for resonators ($g_\mathrm{r}/\omega_\mathrm{r}>0.1$) at  $\varphi_\mathrm{ex}/2\pi\simeq0.5$.

Since the mutual inductance through the rf-SQUID, Eq\erase{s}.~\eqref{eq:Mstar}, is proportional to $\beta\cos{\varphi}/(1+\beta\cos{\varphi})$, where $\beta=2L_\mathrm{sh}/L_\mathrm{J0}$ and $L_0=0$, the coupling strength in the negative direction can be increased relatively easier than in the positive direction at $\varphi/2\pi\simeq0.5$.
Therefore, in our circuit in Fig.~\ref{Design1Circuit}, the geometric mutual inductance $M_0$ is increased by placing the two resonators close each other to increase the coupling strength in the positive direction. 
This allows the coupler to be used in devices that require a large coupling strength in the positive and negative directions, such as those for quantum annealing~\cite{Mukai2019Superconducting,Farhi2000Quantum}.

\begin{figure}[t]
\centering
\includegraphics[width=85mm]{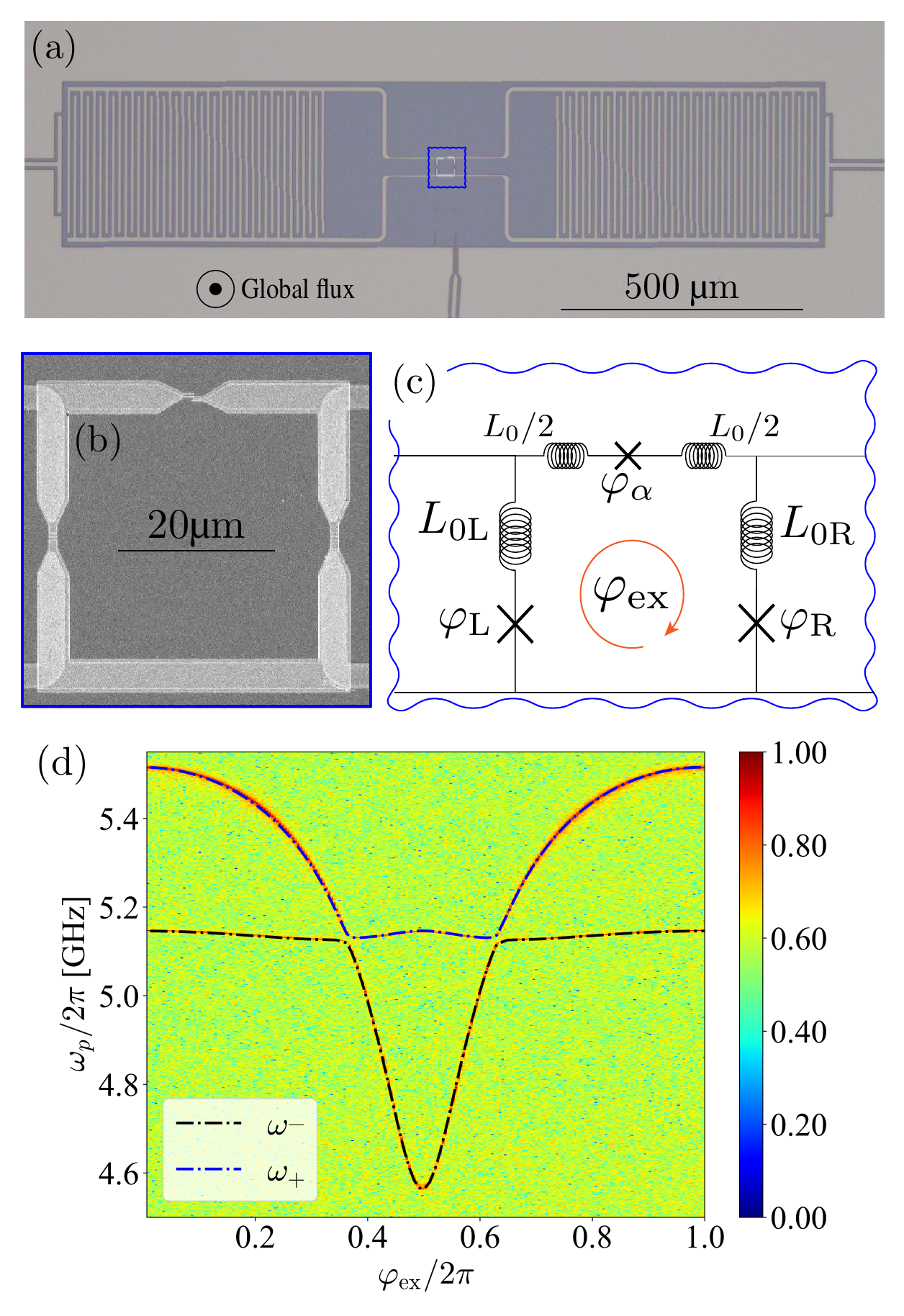}
    \caption{
    (a) Optical microscope image of the coupler circuit with three Josephson junctions.
    (b) SEM image of the coupler area in (a). 
    (c) Schematic of circuit in (a), where $\varphi_\mathrm{ex}$ is controlled by global coil current.
    (d) Measured spectrum of circuit in (a) with fitted curves using parameters    $C_a=485~\si{\femto\farad}$, $C_b=489~\si{\femto\farad}$, $L_{a,b}=1.30~\si{\nano\henry},$ $M_0=34.5~\si{\pico\henry}$, $L_0=137~\si{\pico\henry}$, $L_\mathrm{0L,0R}=33.3~\si{\pico\henry}$, $L_\mathrm{JsL,JsR}=562~\si{\pico\henry}$, and $L_\mathrm{J\alpha}=2.50~\si{\nano\henry}$.
    }
\label{Design2}
\end{figure}

The circuit shown in Fig.~\ref{Design2}, with Josephson junctions instead of $L_\mathrm{sh}$, is another way to increase the coupling strength between resonators.
In the circuit in Fig.~\ref{Design2}, $L_*$ and $M_*$ in Eqs.~\eqref{eq:Mstar} \add{and \eqref{eq:Lstar}} can be written using $L_\mathrm{JsL}/\cos{\varphi_\mathrm{L}}+L_\mathrm{0L}$ and $L_\mathrm{JsR}/\cos{\varphi_\mathrm{R}}+L_\mathrm{0R}$ instead of $L_\mathrm{sh}$ and $L_\mathrm{J}=L_\mathrm{J\alpha}/\cos{\varphi_\alpha}+L_0$.
Each phase for junctions is determined to minimize the potential
\begin{align}
\mathcal{U}_\mathrm{3J} = \qty(\frac{\Phi_0}{2\pi})^2
    &\bigg[
    \frac{1}{2}\frac{\qty(\varphi_\alpha+\varphi_\mathrm{L}+\varphi_\mathrm{R}-\varphi_\mathrm{ext})^2}{L_\mathrm{0L}+L_\mathrm{0R}+L_0} \notag\\
    & 
    -\frac{\cos{\varphi_\alpha}}{L_\mathrm{J\alpha}}
    -\frac{\cos{\varphi_\mathrm{L}}}{L_\mathrm{JsL}}
    -\frac{\cos{\varphi_\mathrm{R}}}{L_\mathrm{JsR}}
    \bigg] \,,
\end{align}
because we consider the SQUID to always be in the ground state.
For the circuit shown in Fig.~\ref{Design2}(a), the measured spectrum with fitted curves using Hamiltonian Eq.~\eqref{eq:H2r} is shown in (d).
This result also indicates that the model [Hamiltonian Eq.~\eqref{eq:H2r}] of circuits with nonlinear inductors reproduce the resonator-coupled system via the rf-SQUID very well.
The tunable range of the coupling constant obtained from the fitting in Fig.~\ref{Design2}(d) is $-291$ to $184~\si{\mega\hertz}$.
In the circuit in Fig.~\ref{Design2}, $L_\mathrm{J\alpha}$ is not designed to be tunable and the screening parameter $\beta=2L_\mathrm{JsL,JsR}/L_\mathrm{J\alpha}\simeq0.45$ is smaller than the maximum value in Fig.~\ref{Design1Spectrum} resulting in a smaller coupling strength in the negative direction.
Also, $M_0$ is much smaller than that of the circuit in Fig.~\ref{Design1Circuit}, and the coupling in the positive direction is not so strong.
In contrast to spectrum of this three-junction coupler in Fig.~\ref{Design2}, one mode is stable while the frequency of the other mode \add{changes}\erase{is changing} in the spectrum in Fig.~\ref{Design1Spectrum}.
This is because the shared inductance between the coupler and the resonator is also modulated \erase{depending on}\add{by} the flux \erase{for}\add{through} the SQUID in the three-junction circuit in Fig.~\ref{Design2}.

\begin{figure*}[t!]
\centering
\includegraphics[width=180mm]{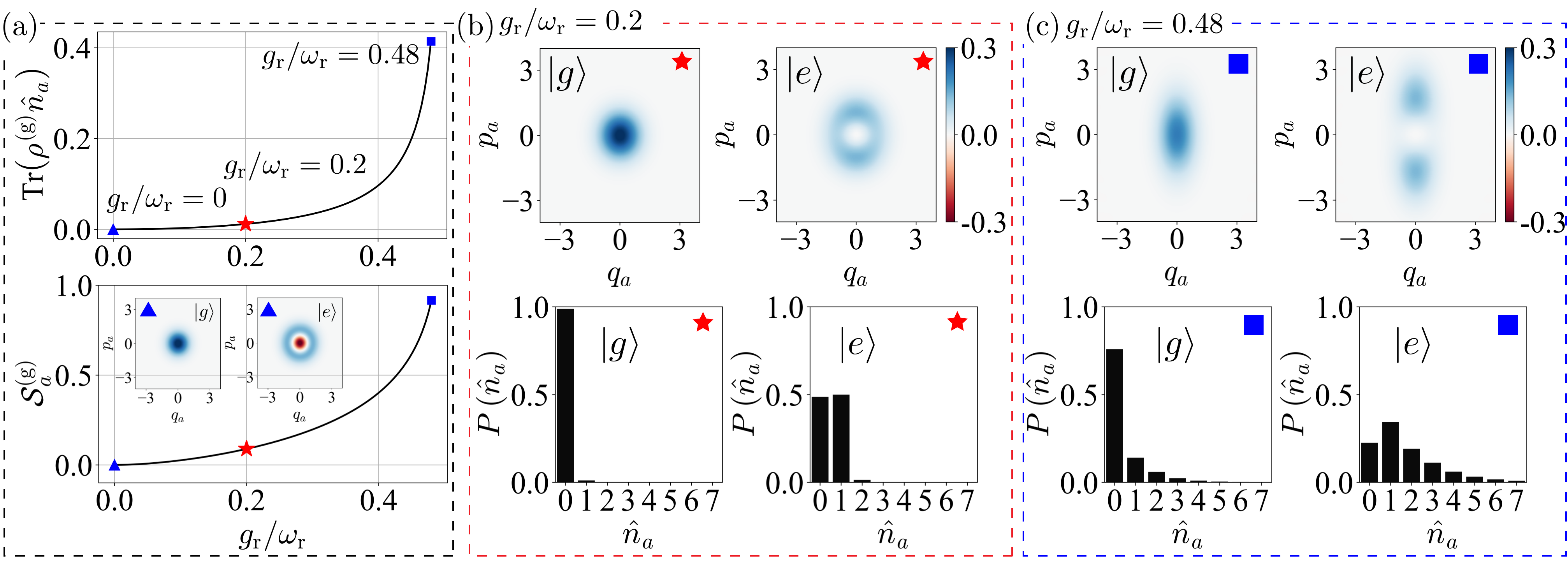}
    \caption{
    (a) Ground-state photon occupation probability and von Neuman entropy as the characteristics of entanglement depending on the energy ratio of the coupling and resonator: $g_\mathrm{r}/\omega_\mathrm{r}$.
    The red star represents the point of maximum coupling in our measured circuit in Fig.~\ref{Design1Circuit}.
    The Wigner functions at $g_\mathrm{r}=0$ in the ground state (left) and first excited state (right) are plotted in the insets. 
    (b) Wigner functions $\mathcal{W}_a(q_a,p_a)$ (up) and Fock distribution $P(\hat{n}_a)$ (bottom) of resonator A with $g_\mathrm{r}/\omega_\mathrm{r}=0.2$. Left and right graphs are for the system [Hamiltonian Eq.~\eqref{eq:H2r}] is in the ground and first excited state ($\ket{g}$ and $\ket{e}$), respectively.
    (c) Same as (b) but with different coupling constant $g_\mathrm{r}/\omega_\mathrm{r}=0.48$, which is almost the limit of the two-resonator coupled system. We use $\omega_a=\omega_b=\omega_\mathrm{r}=5\times 2\pi~\si{\giga\hertz}$ for all numerical calculations in these figures.
    }
\label{USC}
\end{figure*}
\section{Ultrastrong coupling between resonators}\label{sec:USC}
To evaluate the ultrastrong coupling between resonators, we consider the effect of the rotating wave approximation (RWA), which is\erase{valid for} \add{not clearly visible in the spectral measurements in the weak and} the strong coupling regimes.
Applying the RWA to the coupling term in Hamiltonian Eq.~\eqref{eq:H2r}, $\mathcal{H}_\mathrm{int}=-\hbar g_\mathrm{r}(\hat{a}^\dagger-\hat{a})(\hat{b}^\dagger-\hat{b})$, we obtain the simple coupling Hamiltonian $\mathcal{H}_\mathrm{int}^\mathrm{RWA}$ composed of single photon exchange terms and the eigenmodes $\omega_\pm^\mathrm{RWA}$ under this approximation as
\begin{align}
\mathcal{H}_\mathrm{int}^\mathrm{RWA} &=\hbar g_\mathrm{r}(\hat{a}^\dagger \hat{b}+\hat{a}\hat{b}^\dagger) \,, \label{eq:HRWA} \\
\omega_\pm^\mathrm{RWA}&=\frac{\omega_a+\omega_b\pm\sqrt{4g_\mathrm{r}^2+(\omega_a-\omega_b)^2}}{2} \,.
\label{eq:wpmRWA}
\end{align}
\add{When we consider the simple case $\omega_a=\omega_b=\omega_\mathrm{r}$,} the squared frequency difference due to the RWA is written as $\qty(\omega_\pm^\mathrm{RWA})^2-\omega_\pm^2=g_\mathrm{r}^2$\erase{When we consider the simple case $\omega_a=\omega_b=\omega_\mathrm{r}$}, \add{and} the shift (frequency difference) is roughly $0.5g_\mathrm{r}^2/\omega_\pm$.
The approximated modes $\omega_\pm^\mathrm{RWA}$ are plotted in Fig.~\ref{Design1Spectrum}(c) as dashed lines with the same parameter in full Hamiltonian Eq.~\ref{eq:H2r} (Table~\ref{table:SampleData}).
Although $\omega_\pm$ and $\omega_\pm^\mathrm{RWA}$ are close to each other when the coupling constants are close to zero, the shift increases with the coupling constant $g_\mathrm{r}$.
Figure~\ref{Design1Spectrum}(e) shows the obtained RWA shift from the fitting, and the maximum shift is derived as $\omega_-^\mathrm{RWA}-\omega_-=135\times 2\pi~\si{\mega\hertz}$ at $\varphi_\mathrm{ex}/2\pi=0.5$.

Here, \add{as shown in Fig.~\ref{USC}, by numerical simulation, we investigate how the physical properties of the two-resonator-coupled system differ depending on the coupling strength, and the potential of generating a non-classical ground state when the coupling constant increases more than our device.}
\erase{We show the results of numerical calculation of the ultrastrong coupling property in the coupled resonator system.}
First, the mean photon number in the ground state of the system is shown at the top of Fig.~\ref{USC}(a) as a function of the ratio of the coupling constant to the resonator frequency.
It is calculated as the expectation value of $\hat{n}_a=\hat{a}^\dagger\hat{a}$ ($n_a\equiv\expval{\hat{n}_a}$) in the ground-state density matrix $\rho^{(\mathrm{g})}$ of  system Hamiltonian Eq.~\eqref{eq:H2r} with no dissipation, and we obtained $n_a\simeq0.012$ in our coupler ($g_\mathrm{r}/\omega_\mathrm{r}=0.2$).
Second, at the bottom of Fig.~\ref{USC}, we also show the von Neuman entropy of the ground-state reduced density matrix $\rho_a^{(\mathrm{g})}$, which is traced out of resonator B, as $\mathcal{S}_a^{(\mathrm{g})} = -\Tr[ \rho_a^{(\mathrm{g})}\log_2\rho_a^{(\mathrm{g})}]$, and our coupler has $\mathcal{S}_a^{(\mathrm{g})}\simeq0.09$.
As mentioned in section \ref{sec1}, the upper limit of the coupling constant between the resonators is $0.5\omega_\mathrm{r}$, and the mean photon number diverges at $0.5\omega_\mathrm{r}$ in the calculation.
However, in this limit, there is only one resonant mode in the circuit, and we can no longer define two resonators.  
Third, the Wigner function of resonator A is defined as
\begin{equation}
\!\!\!\mathcal{W}_a(q_a,p_a) = \!\frac{1}{2\pi\hbar} \int_{-\infty}^{\infty}
\!\!\!\mel{q_a\!-\frac{x}{2}}{\rho_a}{q_a\!+\frac{x}{2}}e^{ip_a x}\!\dd{x}\,,
\label{eq:wigner}
\end{equation}
where $\hat{a}=q_a+i p_a$, and is shown in Fig.~\ref{USC} for each coupling strength~\cite{Ashhab2010Qubit}.
When there is no coupling, as shown in the insets of Fig.~\ref{USC}(a), we can see the Wigner function of the vacuum state in the ground state (left) and the photon number state of $\hat{n}_a=1$ in the first excited state (right).
In comparison, as the coupling constant increases, the photon distribution and Wigner function in Figs.~\ref{USC}(b) and (c) show squeezed states even in the ground state.
The value at $\hat{n}_a=1$ in the Fock distribution in Fig.~\ref{USC}(b) (left) corresponds to the mean photon number shown by the red star in the top graph of Fig.~\ref{USC}(a).
However, no the obvious difference appear in the Wigner function in $g_\mathrm{r}$ compared with the vacuum state.

\section{Coupling off and crosstalk}\label{sec:cross}

For the coupling devices, the function to turn off the coupling is as important as the ability to set a strong coupling constant for quantum devices.
Undesired interactions between circuit components during the computation process can cause errors and have been studied as a factor that can significantly reduce gate fidelity~\cite{Sung2021Realization}.
Our coupler is designed to be turned off at the closest point of the two observed modes $\omega_\pm$ in the spectrum, and the split width of the two modes $\omega_\pm$ in Eq.~\eqref{eq:wpm} corresponds to the frequency difference between two bare resonators~\cite{noauthor_see_nodate}.

\begin{figure}[h!]
\centering
\includegraphics[width=85mm]{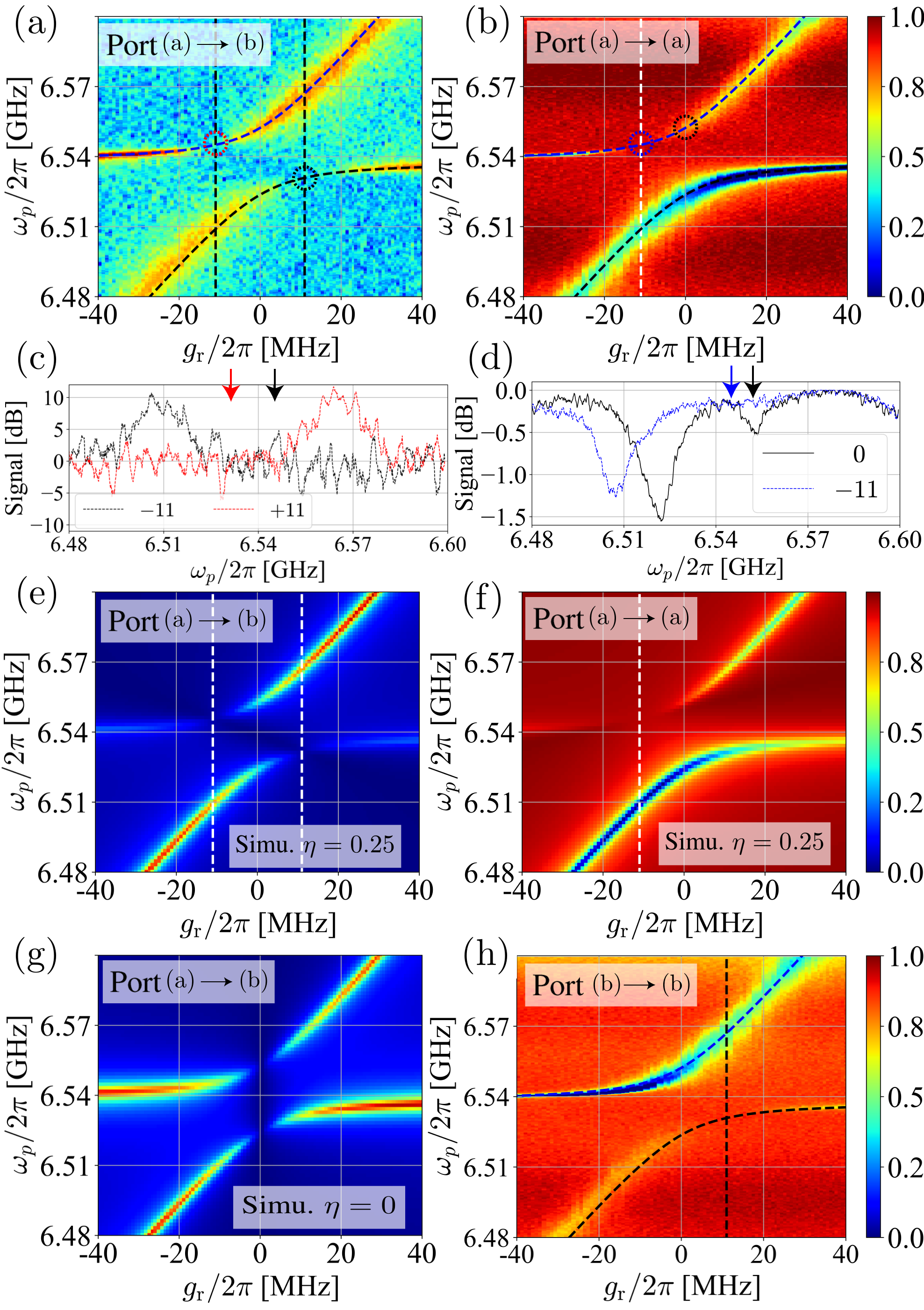}
    \caption{
    (a) Result of transmission spectrum measurement (ports A to B). The spectrum is measured at the enlarged area in Fig.~\ref{Design1Spectrum}(b), where $\varphi_\mathrm{ex}/2\pi<0.5$ and dashed \add{fitted} lines also correspond to those in Fig.~\ref{Design1Spectrum}(b). 
    We take 2 MHz [1 MHz for (b) and (h)] moving average in the frequency space to plot data. Horizontal axis is the coupling constant obtained from fitting [Fig.~\ref{Design1Spectrum}(d)]. 
    (b) Result of reflection spectrum measurement (port A to A) at the same area as (a).
    (c) Cross sections of the spectrum in (a) at $g_\mathrm{r}/2\pi=\pm11~\si{\mega\hertz}$.
    Red and black arrows indicate intersections with fitted curves and $g_\mathrm{r}/2\pi=\pm11~\si{\mega\hertz}$, which are shown in (a) as circles of the same colors.
    (d) Cross section of the spectrum in (b) at $g_\mathrm{r}/2\pi=0$, $-11~\si{\mega\hertz}$.
    Energy absorption cannot be seen (signal disappearance) at the intersection of the fitted curve and the vertical dashed line in (b) [blue arrow in (d)].
    In comparison, we can see the signal at the point of the black arrow, which corresponds to the intersection of the fitted curve and $g_\mathrm{r}=0$. 
    (e)(f) Simulated transmission and reflection spectrum with crosstalk $\eta=0.25$. 
    (h) Result of reflection spectrum measurement (port B to B) at the same area as (a).
    (g) Simulated transmission spectrum with no crosstalk $\eta=0$ at input and output ports.
    (e), (f), and (g) are calculated using the Lindblad master equation with the same parameters as the fitted curve in (a) and $\xi\erase{\sqrt{\kappa_{a,b}}}/2\pi=1.\add{1}\erase{5}~\si{\mega\hertz}$ and $\kappa_{a,b}/2\pi=3.3\times 10^{-3\erase{4}}~\si{\mega\hertz}$.
    }
\label{SpectrumCrossTalk}
\end{figure}

Figure~\ref{SpectrumCrossTalk}(a) shows an enlarged spectrum around the area where $\omega_{\pm}$ is closest ($\varphi_\mathrm{ex}/2\pi<0.5$) in Fig.~\ref{Design1Spectrum}(c)-1.
This spectrum is obtained by measuring the transmission signal from ports A to B in Fig.~\ref{Design1Circuit}(a). 
The signal of $\omega_\pm$ disappears at $g_\mathrm{r}/2\pi\simeq\pm 11~\si{\mega\hertz}$, as can be seen in Fig.~\ref{SpectrumCrossTalk}(a).
If there is no interaction between resonators, no signal from port A can ideally reach port B; thus, the signal should disappear at $g_\mathrm{r}=0$ in the transmission measurement.
However the area where the signal disappears in the measured spectrum is shifted from $g_\mathrm{r}=0$.
Also, for the reflection signals shown in Figs.~\ref{SpectrumCrossTalk}(b) and (h), the signal disappearance points do not correspond to $g_\mathrm{r}=0$
This phenomenon is considered to be caused by crosstalk on the sample.
For example, the input signal from port A reaches resonator B because of the crosstalk through the ground plane of the sample without going through the coupler; then, the transmitted signal can be observed even if $g_\mathrm{r}=0$.

To confirm the effect of crosstalk, we consider a driving Hamiltonian and observables in the spectrum measurement with crosstalk.
After a microwave drive is applied from the input port, we detect the signal emitted to the output port from the whole system as their decays.
Taking into account the crosstalk, the microwave drive Hamiltonian input from port A on a rotational frame with probe frequency $\omega_p$ is described by~\cite{Zheng2017Nonreciprocal,Omelyanchouk2010Quantum}
\add{\begin{equation}
    \mathcal{H}_\mathrm{dA}/\hbar = 
    \qty(1-\eta)\xi(\hat{a}^\dagger+\hat{a})
    +\eta\xi(\hat{b}^\dagger+\hat{b}),
    \label{eq:drive}
\end{equation}}
where $\xi$ is the intensity of the drive \add{and} $\eta$ is the \add{rate}\erase{amount} of crosstalk\erase{, and $\kappa_{a,b}$ is the decay rate of resonators as couplings between each resonator and port}.
Since the left and right (A, B) sides of the circuit are designed to be equal except for the local flux line, we assume that the crosstalk also symmetrically affects each port and resonator.
Even though the two resonators cannot be considered to be completely separate systems except at $g_\mathrm{r}=0$, we assume that each resonator can be treated separately at $g_\mathrm{r}\simeq0$.
Thus, the drive tone with crosstalk from port A would excite resonator A with an intensity of $(1-\eta)\xi$ and resonator B with an intensity of $\eta\xi$.
The measured signal is assumed to be the sum of decays from resonators, and is described as the imaginary part of the annihilation operators $\Im\expval{\hat{a}}$ and $\Im\expval*{\hat{b}}$ from the input output-theory~\cite{Zhao2013Transmission}.
\erase{Because}The crosstalk should also \add{be} take\add{n} into account a signal \add{leakage} from the resonator to the port \add{because the crosstalk can be considered as a coupling constant between port A (B) and resonator B (A).
Therefore, the energy leaks from resonator B to port A with coupling constant $\eta$ when microwaves are applied from port A.
We also assume that the crosstalk between port A and resonator B and the crosstalk between port B and resonator A are equal due to the circuit symmetry (see \ref{sec:A}).
}
\add{Thus, }the transmission coefficient $t_\mathrm{BA}$ from ports A to B and the reflection coefficient $r_\mathrm{AA}$ at port A are represented as
\begin{align}
    t_\mathrm{BA}&=-\frac{\kappa_a
    }{2\xi}\eta\Im\expval{\hat{a}}-\frac{\kappa_b}{2\xi}\qty(1-\eta)\Im\expval*{\hat{b}}  \,,
    \label{eq:tba}\\
    r_\mathrm{AA}&=-\frac{\kappa_a}{2\xi}\qty(1-\eta)\Im\expval{\hat{a}}-\frac{\kappa_b}{2\xi}\eta\Im\expval*{\hat{b}}  \,,
    \label{eq:taa}
\end{align}
\add{where $\kappa_{a,b}$ is the decay rate of resonators.}
\add{And analytical descriptions of Eqs.~\eqref{eq:tba} and \eqref{eq:taa} in the steady state can be written as (see \ref{sec:C})
\begin{widetext}
\begin{align}
    t_\mathrm{BA} &= \frac{\eta(1-\eta)(\kappa_a\kappa_b D_1-2D_2^2)+\qty[\eta^2\kappa_a+(1-\eta)^2\kappa_b]g_\mathrm{r}D_2}{2(D_1^2+D_2^2)} \,,  \label{eq:tba_ana}\\
    r_\mathrm{AA} &= \frac{[\eta^2+(1-\eta)^2]\kappa_a\kappa_b D_1/2
    +[\eta(1-\eta)(\kappa_a+\kappa_b)g_\mathrm{r}
    -\eta^2\kappa_b\delta\omega_a-(1-\eta)^2\kappa_a\delta\omega_b
    ]D_2}{2(D_1^2+D_2^2)}   \label{eq:taa_ana}\,,
\end{align}
\end{widetext}
where, 
$D_1=\delta\omega_a\delta\omega_b-\kappa_a\kappa_b/4-g_\mathrm{r}^2$, $D_2=(\kappa_b\delta\omega_a+\kappa_a\delta\omega_b)/2$, $\delta\omega_k=\omega_k-\omega_\mathrm{d}$, and the drive frequency $\omega_\mathrm{d}$.}
The measured signal shows the absolute values of these coefficients.
In the numerical simulation, we solve the Lindblad-type master equation using the Hamiltonian $\mathcal{H}_\mathrm{2r}+\mathcal{H}_\mathrm{dA}$ with energy decays $\kappa_{a,b}$ to the port, and we obtain the state density matrix of the steady state~\cite{Schlosshauer2007Decoherence}. 
We also use the rotating wave approximation for the coupling term [Eq.~\eqref{eq:HRWA}], because we consider the area where $g_\mathrm{r}$ is small.

The results of calculation using this crosstalk model are shown in Figs.~\ref{SpectrumCrossTalk}(e), (f), and (g). 
Figure~\ref{SpectrumCrossTalk}(e) shows the simulation result of the transmission signal $t_\mathrm{BA}$ in the case of crosstalk $\eta=0.25$, and the signal disappears at $g_\mathrm{r}/2\pi\simeq\pm11~\si{\mega\hertz}$, which corresponds well to the measurement result in (a).
In contrast, Fig.~\ref{SpectrumCrossTalk}(g) shows the case of non-crosstalk ($\eta=0$), and the transmission signal disappears at point $g_\mathrm{r}=0$.
The reason why the signal at $g_\mathrm{r}=0$ is observed in the experiment is that the two uncoupled resonators are both excited via the crosstalk and the signal leaks to both ports via the crosstalk.
In the case of the reflection simulation in Fig.~\ref{SpectrumCrossTalk}(f), the $\omega_-$ signal is still visible in (b), while the $\omega_+$ signal disappears at $g_\mathrm{r}/2\pi\simeq-11\si{\mega\hertz}$.
If there is no interaction between resonators in reflection measurement, $\omega_-$ can always be seen because $\omega_-$ corresponds to the bare frequency of resonator A $\omega_{a\add{0}}$ at $g_\mathrm{r}=0$ ($\omega_{a0}<\omega_{b0}$); these results are also consistent with the experimental result in (b).
Contrarily, in the \add{reflection}\erase{refraction} measurement at port B, $\omega_+$ can always be seen in Fig.~\ref{SpectrumCrossTalk}(h).
The signal disappearance in the presence of crosstalk is caused by destructive interference between $\expval{\hat{a}}$ and $\expval*{\hat{b}}$, where the amplitudes of the microwave signals canceled each other out. 
This is because the phases of $\expval{\hat{a}}$ and $\expval*{\hat{b}}$ are determined by the coupling constant $g_\mathrm{r}$ and the crosstalk $\eta$.
A similar shift of the signal disappearance point can be seen in the spectrum in Ref.~\onlinecite{Wulschner2016Tunable}, which also seems to be the same phenomenon.
\add{The amount of crosstalk is determined by the geometry of circuit components and ground stability.
If we couple qubits with each resonator, qubit--qubit crosstalk is expected to be much smaller than $\eta=0.25$. This is because the qubit size is much smaller than the port and the resonator.
}

\section{Conclusion}
We implemented and evaluated the ultrastrong tunable coupler between resonators using an rf-SQUID.
Our circuit model treating junctions as tunable inductances reproduces the experimental results very well.
The fitting of the spectrum shows the high tunablity of the coupling constant $g_\mathrm{r}/2\pi$ from $-1086$ $\si{\mega\hertz}$ (antiferromagnetic) to 604 $\si{\mega\hertz}$ (ferromagnetic).
Thus, our coupler achieves an ultrastrong coupling regime, and we observe a breaking of the rotational wave approximation in the spectrum measurement.
Turning off the coupling is an important function to the construction of quantum devices.
By comparison with crosstalk simulations, we confirms the existence of a zero-coupling point.
\add{We assume that this ultrastrong coupler is compatible not only with lumped element resonators but also with other shape of resonators such as coplanar waveguide resonator.}
On the basis of these results, our coupler could be used in a full coupling annealer, as well as in NISQ devices and quantum simulators that require more dense connections and/or coupling qubits far apart.
It is also expected to be applied to the up- and down-conversions devices of photons or to research on quantum phenomena using ultrastrong coupling such as the generation of entanglement states.

\section{Acknowledgement}
We thank S. Watabe, Y. Matsuzaki, T. Nikuni, Y. Zhou, S. Shirai, R. Wang, S. Kwon, Y. Hashizume, S. Kawabata, and T. Yoshioka for their thoughtful comments on this research. We also thank K. Kusuyama, K. Nittoh, and L. Szikszai for their support during sample fabrication. 
In sections \ref{sec:USC} and \ref{sec:cross}, we partly used functions in the quantum toolbox in Python (QuTip)~\cite{Johansson2012QuTiP,Johansson2013QuTiP}.
This paper was based on results obtained from a project, JPNP16007, commissioned by
the New Energy and Industrial Technology Development Organization (NEDO), Japan. Supporting from JST CREST (Grant No. JPMJCR1676) and Moonshot R \& D (Grant No. JPMJMS2067) is also appreciated.

\bibliography{UScoupler-20210729.bib} 

\begin{thebibliography}{49}
\bibitem [{noa()}]{noauthor_see_nodate}%
  \BibitemOpen
  \href@noop {} {}\bibinfo {note} {In the spectrum, the point where $\omega_-$
  and $\omega_+$ are the closest to each other corresponds to the point where
  the coupling constant becomes zero. This is because in the case of a qubit
  and a resonator, the horizontal axis is usually the energy of the qubit, but
  in the spectrum of the coupler, the horizontal axis corresponds to the
  coupling constant, so the larger the $g_\mathrm{r}$ is, the farther apart the
  two modes $\omega_-$ and $\omega_+$ become.}\BibitemShut {Stop}%
\end{thebibliography}
\appendix
\renewcommand{\thesection}{Appendix \Alph{section}}
\renewcommand{\theequation}{\Alph{section}\arabic{equation}}
\renewcommand{\thefigure}{\Alph{section}\arabic{figure}}
\setcounter{figure}{0}

\section{\add{Observable eigenmodes $\omega_\pm$}}\label{sec:A0}
\add{
Here, we approximate $\omega_a=\omega_b$ ($C_a=C_b$ and $L_a=L_b$) and $L_0=0$, because we equally designed resonator A and B, and $L_0\ll L_\mathrm{J0}$.
In this case, two eigenmodes $\omega_\pm$ [Eq.~\eqref{eq:wpm}] can be described as
\begin{align}
    \text{the case $g_\mathrm{r}\geq0$}&\,, \notag\\
    \omega_+&=1/\sqrt{(L_a+L_*)C_a} \,,\\
    \omega_-&=1/\sqrt{(L_a+L_\mathrm{sh}+2M_0)C_a}\,,\\
    \text{the case $g_\mathrm{r}<0$}&\,,\notag \\
    \omega_+&=1/\sqrt{(L_a+L_\mathrm{sh}+2M_0)C_a} \,,\\
    \omega_-&=1/\sqrt{(L_a+L_*)C_a}\,.
\end{align}
Also, when $g_\mathrm{r}=0$, $L_*=L_\mathrm{sh}+2M_0$.
In above equations, only $L_*$ has the flux bias dependence, thus one of the eigenmodes has no flux modulation while the other eigenmode has flux modulation in the spectrum.
}

\section{Measurement environment}\label{sec:A}
In the measurement setup, visible frequency range is in 4 to 8 GHz due to the cryogenic microwave components including bandpass filter, the cryogenic amplifier, and circulator. The coil current 115 $\si{\micro}$A is the bias point to give the measurable maximum coupling constant in our setup.
Moreover, the local flux bias also affect to the dc-SQUID loop. The ratio of rf-SQUID loop flux to dc-SQUID loop flux by current from the local flux bias line is 313 which is included in the fitting function as a constant value but it is actually negligibly small effect for the spectrum.

All measurements were performed in a dilution refrigerator below $10~\si{\milli\kelvin}$, and the input power at the port of the sample holder is around -130 dBm. 

\begin{figure}[h]
\centering
\includegraphics[width=85mm]{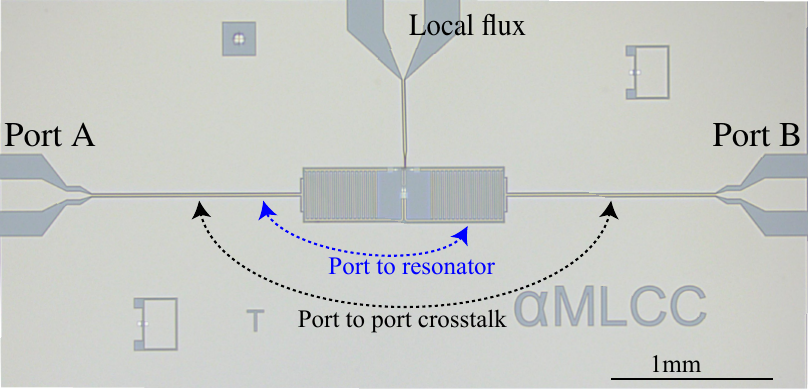}
    \caption{
    \add{Chip image of Fig.~\ref{Design1Circuit}(a). Arrows indicate interaction due to crosstalk.}}
\label{Chip}
\end{figure}

\add{The chip image of Fig.~\ref{Design1Circuit}(a) is shown in Fig.~\ref{Chip}.
The crosstalk exists between ports and between a port and a resonator in our sample through environment, especially the ground plane. 
}
\setcounter{figure}{0}
\begin{figure*}[t!]
\centering
\includegraphics[width=180mm]{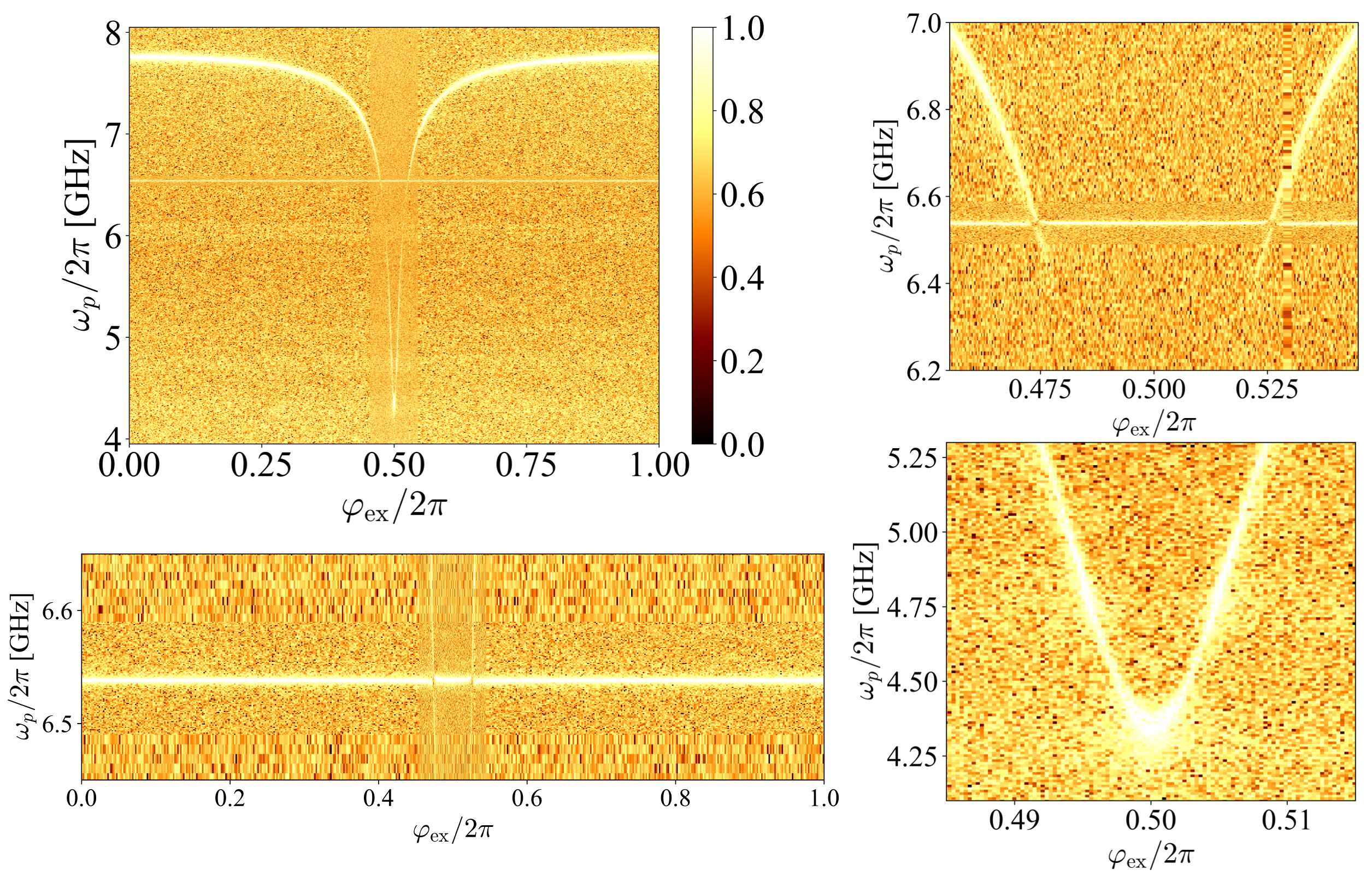}
    \caption{
    \add{Measured raw spectrum data same as Fig.~\ref{Design1Spectrum}(b), (c)-1, (c)-2, and (c)-3 without fitted curves.}
    }
\label{Raw}
\end{figure*}

\section{Spectrum fitting}\label{sec:B}
In Fig.~\ref{Design1Circuit} of main text, the Josephson junction is treated as an tunable inductance, but the superconducting-insulator-superconducting junction actually has capacitance and resistance, which are usually described by the resistively-capacitively-shunted-Junction (RCSJ) model.
In the RCSJ model, a junction are assumed to be a parallel LCR circuit with their plasma frequency $\omega_\mathrm{J0}=1/\sqrt{L_\mathrm{J0}C_\mathrm{J}}$.
The impedance of the parallel LCR circuit $ Z_\mathrm{JJ}(\omega)$ is given by the following equation depending on the frequency $\omega$ of the microwave entering the junction:
\begin{align}
    Z_\mathrm{JJ}(\omega)
    &=\qty(\frac{1}{i\omega L_\mathrm{J}}+i\omega C_\mathrm{J}+\frac{1}{R_\mathrm{J}})^{-1} \notag\\
    &=i\omega L_\mathrm{J0}\qty(\cos{\varphi}-\frac{\omega^2}{\omega_\mathrm{J0}^2}+\frac{i\omega L_\mathrm{J0}}{R})^{-1}\notag \\
    &=\frac{Z_L}{\cos{\varphi}-\gamma}
    \,,
    \label{eq:z}
\end{align}
where $Z_L\equiv i\omega L_\mathrm{J0}$, $\gamma\equiv\bar{\omega}^2/\omega_\mathrm{J0}^2$, and the resistance R is ignored in the fitting ($R\rightarrow\infty$).
Thereby, the effect of capacitance in the fitting function in the main text can be considered as an offset $\gamma$ of the cosine modulation of inductance.
Although the probing microwave frequency is not constant in the spectrum, we take it to be an average, and $\gamma=0.053$ is obtained from the fitting parameter in Fig.~\ref{Design1Spectrum}.

The resistance in Eq.~\eqref{eq:z} has a property to increase the minimum frequency of $\omega_-$ in Fig.~\ref{Design1Spectrum}(c)-3, corresponding to reduce the coupling constant $g_\mathrm{r}$.
Since $M_*=\beta\cos{\varphi}/(1+\beta\cos{\varphi})$, when $\beta=1$, the coupling constant $g_\mathrm{r}$ diverges at $\varphi/2\pi=0.5$. 
But the imaginary part in Eq.~\eqref{eq:z} prevents $g_\mathrm{r}$ to diverge.
Similarly, dissipation in resonators and SQUIDs also prevent $g_\mathrm{r}$ to diverge.
However, the junction's dissipation [$R$ in Eq.~\eqref{eq:z}] itself does not need to be introduce in the fitting function to reproduce the measured spectrum.

\add{The raw data of the spectrum Fig.~\ref{Design1Spectrum}(b) and (c) before fitting are shown in Fig.~\ref{Raw}}

\add{
\section{Heisenberg equation}\label{sec:C}
Equations~\eqref{eq:tba_ana} and \eqref{eq:taa_ana} are derived from the Heisenberg equation using system Hamiltonian $\mathcal{H}_\mathrm{sys}$ on rotating frame with RWA;
\begin{align}
    \mathcal{H}_\mathrm{sys}/\hbar=&\,\delta\omega_a\hat{a}^\dagger\hat{a}
    +\delta\omega_b\hat{b}^\dagger\hat{b}
    +g_\mathrm{r}(\hat{a}^\dagger\hat{b}+\hat{a}\hat{b}^\dagger) \notag \\
    &+(1-\eta)\xi(\hat{a}^\dagger+\hat{a})+\eta\xi(\hat{b}+\hat{b}^\dagger)
    \label{eq:Hsys}
\end{align}
The driving term of above Hamiltonian include crosstalk when the drive tone is applied from port A.
Then, Heisenberg equations about $\hat{a}$ and $\hat{b}$ in Hamiltonian Eq.~\eqref{eq:Hsys} are described by
\begin{align}
    \frac{\mathrm{d}\hat{a}}{\mathrm{d}t}=&-i\delta\omega_a\hat{a}-ig_\mathrm{r}\hat{b}-i(1-\eta)\xi-\frac{\kappa_a}{2}\hat{a} \,,\label{eq:HEa}\\
    \frac{\mathrm{d}\hat{b}}{\mathrm{d}t}=&-i\delta\omega_b\hat{b}-ig_\mathrm{r}\hat{a}-i\eta\xi-\frac{\kappa_b}{2}\hat{b} \,.\label{eq:HEb}
\end{align}
In the spectrum measurement, we consider the steady state $\dv*{\hat{a}}{t}=0$, and $\mathrm{d}\hat{b}/\mathrm{d}t=0$, and transmission and reflection coefficients Eqs.~\eqref{eq:tba_ana} and \eqref{eq:taa_ana} are derived from Eqs.~\eqref{eq:tba}, \eqref{eq:taa}, \eqref{eq:HEa}, and \eqref{eq:HEb}.
}

\add{
In the transmission coefficient $t_\mathrm{BA}$ [Eq.~\eqref{eq:tba_ana}], if there is no crosstalk ($\eta=0$), transmission signal disappears ($t_\mathrm{BA}=0$) at $g_\mathrm{r}=0$.
However, in Fig.~\ref{sec:cross}, when there is a finite crosstalk, the signal disappears at
\begin{align}
    g_\mathrm{r}=\pm\frac{\eta(1-\eta)}{\eta^2+(1-\eta)^2}(\omega_+^\mathrm{RWA}-\omega_-^\mathrm{RWA}) \,,
\end{align}
where $\kappa_a=\kappa_b$, and $D_1\simeq \delta\omega_a\delta\omega_b-g_\mathrm{r}^2$ for $\kappa_a\ll g_\mathrm{r}$ because $\kappa_a$ is \si{\kilo\hertz} and $g_\mathrm{r}$ is much large around \si{\mega\hertz}.
}




\end{document}